\documentclass[11pt,letterpaper]{article}
\usepackage[dvips]{epsfig}
\usepackage{epsf,amsfonts,amssymb}

\newcommand{\be}{\begin{equation}}
\newcommand{\ee}{\end{equation}}
\newcommand{\ben}{\begin{displaymath}}
\newcommand{\een}{\end{displaymath}}
\newcommand{\bea}{\begin{eqnarray}}
\newcommand{\eea}{\end{eqnarray}}
\newcommand{\bean}{\begin{eqnarray*}}
\newcommand{\eean}{\end{eqnarray*}}

\newcommand{\ba}{\begin{array}}
\newcommand{\ea}{\end{array}}
\newcommand{\bi}{\begin{itemize}}
\newcommand{\ei}{\end{itemize}}

\renewcommand{\O}{\ensuremath{\Omega}}

\newcommand{\gsim}{\mathrel{\raisebox{-.6ex}{$\stackrel{\textstyle>}{\sim}$}}}
\newcommand{\lsim}{\mathrel{\raisebox{-.6ex}{$\stackrel{\textstyle<}{\sim}$}}}

\newcommand{\eg}{{\it e.g.,}\ }
\newcommand{\ie}{{\it i.e.,}\ }

\newcommand{\ymax}{y_{\rm max}}

\newcommand{\beq}{\begin{equation}}
\newcommand{\eeq}{\end{equation}}
\newcommand{\beqr}{\begin{displaymath}}
\newcommand{\eeqr}{\end{displaymath}}
\newcommand{\beqa}{\begin{eqnarray}}
\newcommand{\eeqa}{\end{eqnarray}}
\newcommand{\beqar}{\begin{eqnarray*}}
\newcommand{\eeqar}{\end{eqnarray*}}
\newcommand{\m}{\ensuremath{\mu}}
\newcommand{\n}{\ensuremath{\nu}}
\newcommand{\p}{\ensuremath{\pi}}

\newcommand{\h}{\ensuremath{\eta}}

\newcommand{\reef}[1]{(\ref{#1})}

\newcommand{\rp}{r_{+}}
\renewcommand{\to}{\ensuremath{\tilde{\omega}}}
\newcommand{\tO}{\ensuremath{\tilde{\Omega}}}
\newcommand{\tm}{\ensuremath{\tilde{\mu}}}
\newcommand{\tk}{\ensuremath{\tilde{k}}}

\newcommand{\kmax}{k_{\mathrm{max}}}
\newcommand{\tkmax}{\tilde{k}_{\mathrm{max}}}

\newcommand{\model}{\mathrm{model}}

\begin{document}

\setlength{\unitlength}{1mm}

\thispagestyle{empty}
\vspace*{3cm}

\begin{center}
{\bf \Large Black Rings, Boosted Strings}\\
{\bf \Large and Gregory-Laflamme}\vspace*{2cm}

{\bf Jordan L. Hovdebo}\footnote{E-mail: {\tt
jlhovdeb@sciborg.uwaterloo.ca}} {\bf and Robert C.
Myers}\footnote{E-mail: {\tt rmyers@perimeterinstitute.ca}}

\vspace*{0.2cm}

{\it Perimeter Institute for Theoretical Physics}\\
{\it 31 Caroline Street North, Waterloo, Ontario N2L 2Y5, Canada}\\[.5em]

{\it Department of Physics, University of Waterloo}\\
{\it Waterloo, Ontario N2L 3G1, Canada}\\[.5em]

\vspace{2cm} {\bf ABSTRACT} 
\end{center}
We investigate the Gregory-Laflamme instability for black strings
carrying KK-momentum along the internal direction. We demonstrate a
simple kinematical relation between the thresholds of the classical
instability for the boosted and static black strings. We also find
that Sorkin's critical dimension depends on the internal velocity
and in fact disappears for sufficiently large boosts. Our analysis
implies the existence of an analogous instability for the
five-dimensional black ring of Emparan and Reall. We also use our
results for boosted black strings to construct a simple  model of
the black ring and argue that such rings exist in any number of
space-time dimensions.

\vfill \setcounter{page}{0} \setcounter{footnote}{0}
\newpage

\section{Introduction}

After many years of investigation, higher dimensional general
relativity still continues to be a rich source of new ideas and
physics. It is now eighty years ago that Kaluza \cite{kaluza} and
Klein \cite{klein} first entertained the idea of general relativity
in higher dimensions as a route towards the unification of gravity
with the other forces in nature. Their nascent explorations laid
foundations for modern superstring and M-theory. The study of higher
dimensional general relativity provides important insights into the
structure of these theories. Further in recent braneworld scenarios,
the extra dimensions are much larger than the Planck scale and so
the study of classical Einstein equations in higher dimensions is
necessary to understand the phenomenology of these models.

While there has been a great deal of activity studying  ``black
objects'' in higher dimensions, particularly in string theory
\cite{newsols}, there is clear evidence that our four-dimensional
intuition leads us astray in thinking about the physics of event
horizons in higher dimensional gravity. For example, an interesting
corollary of the early theoretical investigations of black holes in
four-dimensions was that each connected component of a stationary
horizon must have the topology of a two-sphere \cite{round}.
However, this result is easily evaded in higher dimensions. As a
simple example, consider the four-dimensional Schwarschild metric
combined with a flat metric on $R^m$. This space-time is an extended
black hole solution of Einstein's equations in 4+$m$ dimensions, and
the topology of the horizon is $S^2\times R^m$. Clearly, this
straightforward construction is easily extended to constructing many
other higher-dimensional black holes whose horizons inherit the
topology of the ``appended'' manifold.\footnote{Similar solutions
arise for four dimensions in the presence of a negative cosmological
constant \cite{neg}.} These solutions describe extended objects in
that the geometry is not asymptotically flat in all 3+$m$ spatial
directions and so one might have conjectured that all localized
black objects would have a spherical horizon. However, this hope was
eliminated by Emparan and Reall \cite{ring} who constructed an
explicit five-dimensional metric describing a black ring with
horizon topology $S^2\times S^1$. The circle direction in these
solutions is supported against collapse by angular momentum carried
in this direction, as was anticipated much earlier in \cite{MP}.

These black ring solutions also eliminated any possibility of
extending the usual black hole uniqueness theorems beyond four
dimensions. In four-dimensional general relativity, work on black
hole uniqueness theorems began with the pioneering work of
Israel \cite{firsthair}. The no-hair results are now rigorously
established for Einstein gravity coupled to Maxwell fields and
various other simple matter systems \cite{hair}. While in string
theory, we study more complicated matter field couplings (as well as
spacetime dimensions beyond four), the plethora of new
solutions \cite{newsols} still respected the spirit of the no-hair
theorems in that the black hole geometries are still completely
determined by some small set of charges. However, the black rings
\cite{ring} explicitly provided two solutions for which the mass and
spin were degenerate with five-dimensional spinning black holes
\cite{MP}. This non-uniqueness was further extended to a continuous
degeneracy with the introduction of dipole charges
\cite{emparan-dipole}.

One open question is whether or not such black rings exist in more
than five dimensions.  One argument suggesting that five dimensions
is special comes from considering the scaling of the Newtonian
gravitational and centripetal forces. In this sense, five-dimensions
is unique in that it is only for $D=5$ that these forces scale in
the same way and can be stably balanced. Of course, this is purely a
classical argument which need not be true in the fully relativistic
theory and further it ignores the tension of the ring. It is part of
the goal of this paper to address this question.

In considering spinning black holes and rings, four dimensions is
also distinguished from higher dimensions by the Kerr bound. While
there is an upper bound on the angular momentum per unit mass of a
four-dimensional black hole, no such bound exists for black holes in
dimensions higher than five \cite{MP}. The five-dimensional black
rings also remove this bound in higher dimensions \cite{ring}.

Even more strikingly, in contrast to the stability theorems proven
for four-dimensional black holes \cite{stable4}, Gregory and
Laflamme \cite{gla1,gla2} have shown that extended black branes are
unstable.  The spectrum of metric perturbations contains a growing
mode that causes a ripple in the apparent horizon. The endpoint of
the instability is not completely clear, however, a fascinating
picture is emerging \cite{kolrev}. Interestingly, it was shown in
\cite{ultra} the Gregory-Laflamme instability dynamically enforces
the ``Kerr bound'' for $D \ge 6$. Perhaps a stability criterion will
restore some of the restrictions which are seen to apply to black
holes in four dimensions.

In the present paper, we investigate the Gregory-Laflamme
instability for black strings carrying Kaluza-Klein (KK) momentum.
These solutions are easily constructed by boosting the static black
string metrics. We begin in Section
\ref{sec:gregory_laflamme_instability} with a review of the
Gregory-Laflamme instability for static black strings. The
discussion of boosted black strings begins in Section
\ref{sec:kksolutions}, where we first present the solutions carrying
KK-momentum and then consider their stability with global
thermodynamic arguments. We then adapt the usual numerical analysis
of the Gregory-Laflamme instability to these boosted solutions. We
demonstrate a simple kinematical relation between the thresholds of
the instability for boosted and static black strings with a fixed
horizon radius. Comparing the numerical results with the previous
global analysis, we find that Sorkin's critical dimension \cite{sor}
depends on the boost velocity. In section \ref{sec:loop_of_string},
we apply our results to a discussion of the stability of the black
ring solutions of Emparan and Reall \cite{ring}. As already
anticipated there, we find that large black rings will suffer from a
Gregory-Laflamme instability. Our analysis allows us to argue that
black rings will exist in any dimension higher than five as well.


\section{Gregory-Laflamme Instability}
\label{sec:gregory_laflamme_instability}

The detailed calculation of the instability of the boosted black
strings will be an extension of the original analysis of Gregory and
Laflamme \cite{gla1,gla2}. Hence we begin here by reviewing the
stability analysis for static black strings.\footnote{Note, however,
that our gauge fixing follows \cite{ammw} which differs from that in
the original analysis of \cite{gla1,gla2}. The present gauge fixing
\cite{ammw} has the advantages that it succeeds in completely fixing
the gauge and it is well-behaved in the limit of vanishing $k$.} For
the static string in $D=n+4$ dimensions, the background metric can
be written as
\begin{equation}
ds^2=-f(r)\,dt^2+\frac{dr^2}{f(r)}+r^2 d\O^{\,2}_{n+1}+dz^2 \ ,
\label{static}
\end{equation}
where $d\Omega_{n+1}^{\,2}$ is the metric on a unit $(n+1)$-sphere
and
\begin{equation}
f(r) = 1-\left ( \frac{r_+}{r} \right )^{n} \ .
\end{equation}
The event horizon is situated at $r=\rp$ and we imagine that the
$z$ direction is periodically identified with $z = z + 2\pi R$.

Now we seek to solve the linearized Einstein equations for
perturbations around the above background \reef{static}. The full
metric is written as
\begin{equation}
g_{\mu \nu} = \tilde{g}_{\mu \nu} + h_{\mu \nu} \ ,
\end{equation}
where $\tilde{g}_{\mu \nu}$ is the background metric \reef{static}
and $h_{\mu \nu}$ is the small perturbation. We will restrict the
stability analysis to the S-wave sector on the ($n+1$)-sphere as
it can be proven that modes with $\ell\ne0$ are all completely
stable. This is apparent following the line of argument originally
presented in \cite{reall}.  The threshold for any instability
should correspond to a time-independent mode. This mode can be
analytically continued to a negative mode of the Euclidean
Schwarzschild solution, however, Gross, Perry and Yaffe \cite{gpy}
have shown that the existence of such a mode is unique to the
s-wave sector. Hence we write the perturbations as
\begin{equation}
h_{\mu \nu} = {\mathcal Re} \left [ e^{\Omega\, t + i k\, z}
a_{\mu \nu}(r) \right ] \ , \label{eq:ansatz}
\end{equation}
where $\Omega$ and $k$ are assumed to be real and $a_{\mu \nu}$ is
chosen to respect the spherical symmetry, \eg $a_{z\theta}=0$.
Hence solutions with $\Omega> 0$ correspond to instabilities of
the static black string. The above ansatz \reef{eq:ansatz} can be
further simplified with infinitesimal diffeomorphisms. Using a
diffeomorphism with the same $t$ and $z$ dependence as above, the
perturbation may be reduced to a form where the only nonvanishing
components of $a_{\mu\nu}$ are:
\begin{eqnarray}
&&a_{tt} = h_t(r)\ ,\quad a_{rr} = h_r(r)\ , \quad a_{zz}= h_z(r)
\ ,\nonumber\\
&&\qquad a_{tr} = \Omega\, h_v(r)\ ,\quad a_{zr} = - i k\, h_v(r)
\ . \label{rprof}
\end{eqnarray}
Note that even though $a_{\theta\theta}=0=a_{\phi\phi}$, these
perturbations can cause rippling in the position of the {\it
apparent} horizon along the internal direction \cite{gla2}.

The linearized Einstein equations give a set of coupled equations
determining the four radial profiles above. However, we may
eliminate $h_v,h_r$ and $h_t$ from these equations to produce a
single second order equation for $h_z$:
\begin{eqnarray}
h_z''(r) & + & p(r)\, h_z'(r)\ +\ q(r)\, h_z(r) = \Omega^2 w(r)
h_z(r) \label{eq:line}\\ \cr
 p(r) & = & \frac{1}{r} \left( 1 +
\frac{n}{f(r)} - \frac{4 (2+n) \, k^2 r^2}{2 \, k^2 r^2 + n (1+n)
\left( \frac{r_+}{r} \right)^n} \right)  \cr q(r) & = &
\frac{1}{r^2} \left( - \frac{k^2 r^2}{f(r)} \; \frac{2 \, k^2 r^2
- n(3+n) \left(\frac{r_+}{r}\right)^n}{2 \, k^2 r^2 + n(1+n)
 \left(\frac{r_+}{r}\right)^n} \right) \cr
w(r) & = & \frac{1}{f(r)^2} \label{eq:hyeqn}
\end{eqnarray}

Next we must determine the appropriate boundary conditions on
$h_z(r)$ at the horizon and asymptotic infinity for a physical
solution. First near the horizon, the radial equation
\reef{eq:line} simplifies considerably yielding solutions
\begin{equation}
h_z = A e^{\Omega\, r_*} + B e^{-\Omega\, r_*} \ .
\label{eq:horizonbc}
\end{equation}
Here $r_*$ is the tortoise coordinate defined by $dr_*/dr=1/f$ and
with which the horizon appears at $r_*\rightarrow-\infty$. Now in
principle, we would choose initial data for the perturbation on a
Cauchy surface extending to the future horizon and demand that the
perturbation be finite there. Hence we require that $B=0$ for
physical solutions.\footnote{While the present argument is
somewhat superficial, a more careful treatment yields the same
result \cite{gla1,gla2}.}

Eq.~\reef{eq:line} also simplifies as $r \rightarrow \infty$. The
asymptotic solutions behave differently depending on whether $n=1$
or $n\ge2$.  For $n=1$, the regular solutions take the form
\begin{equation}
h_z \sim e^{- \mu\,r} r^{2 - \frac{\Omega^2+\mu^2}{2 \mu}r_+}\ ,
\label{eq:asymbc5d}
\end{equation}
where $\mu^2\equiv\Omega^2+k^2$. For $n\ge2$, they are
\begin{equation}
h_z \sim e^{- \mu\, r} r^{\frac{n+3}{2}} \ , \label{eq:asymbc}
\end{equation}
with the same definition for $\mu$. Hence we expect that the
unstable perturbations are localized near the horizon with a
characteristic size $\mu^{-1}$.

\begin{figure}[h]
\begin{center}
\resizebox{10cm}{5cm}{\includegraphics{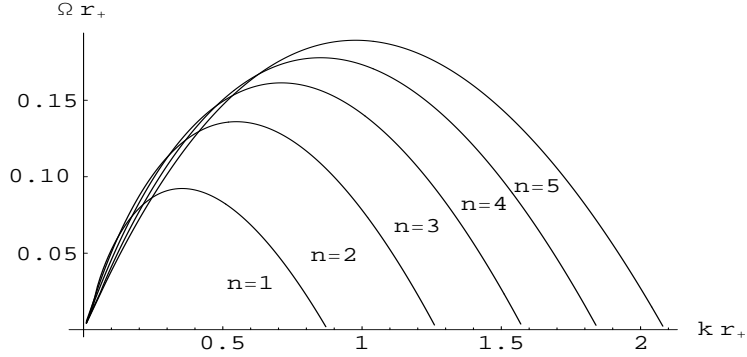}}
\caption{Unstable frequencies and wavenumbers for the static black
string.} \label{fig:Omega_vs_k_no_boost}
\end{center}
\end{figure}

The instabilities can be determined as follows: For a fixed value
of $k$, we choose $\Omega$ and set the asymptotic conditions
according to eq.~(\ref{eq:asymbc5d}) or (\ref{eq:asymbc}). The
radial equation \reef{eq:line} is integrated in numerically to $r
\approx r_+$. Here we match the numerical solution to the
near-horizon solution (\ref{eq:horizonbc}) which determines
the ratio $B/A$ for the chosen value of $\Omega$. By
varying $\Omega$, we may tune this ratio to satisfy the physical
boundary condition at the horizon, \ie $B=0$. We find solutions
for a range of $k$ from $0$ up to a maximum value
$k_{\mathrm{max}}$. Figure \ref{fig:Omega_vs_k_no_boost} shows the
resulting solutions for various spacetime dimensions. The critical
value $k_{\mathrm{max}}$ corresponds to the threshold of the
Gregory-Laflamme instability and is set by the only dimensionful
parameter in the background, $\rp$, up to a factor of order one.
Table \ref{tab:unboosted-coeffs} tabulates $k_{\mathrm{max}}$ for
different values of $n$.
\begin{table}
\begin{center}
\begin{tabular}{|c|c|c|c|c|c|}
\hline
n & 1 & 2 & 3 & 4 & 5 \\
\hline
$k_{\mathrm{max}}\rp$ & $0.876$ & $1.269$ & $1.581$ & $1.849$ & $2.087$ \\
\hline
\end{tabular}
\caption{Maximum wavenumber corresponding to the marginally
unstable mode of the static black string in various dimensions
$D=n+4$.} \label{tab:unboosted-coeffs}
\end{center}
\end{table}

When the coordinate along the string is periodic, the allowed
values of $k$ are discrete, \ie for $z=z+2\pi R$, $k= n/R$ with
$n$ an integer. Hence for small $R$, the system is stable when
$\kmax\ge 1/R$. However, for $R>1/\kmax$, the lowest wavenumber,
allowed by periodicity, falls in the unstable range and the black
string is unstable.


\section{Boosted Black Strings} \label{sec:kksolutions}

Our focus at present is ``boosted black
strings,'' \ie stationary black string solutions carrying momentum
along their length. Such solutions can be obtained by simply
boosting the static solution \reef{static} along the $z$ direction
\begin{equation}
ds^2=-dt^2+\frac{dr^2}{f(r)}+r^2 d\O^{\,2}_{n+1}+dz^2  +
(1-f)\cosh^2 \beta \left (dt + \tanh \beta ~ dz \right )^2 \ ,
\label{eq:metric}
\end{equation}
where the boost velocity is given by $v=\tanh\beta$, and as before
\begin{equation}
f(r) = 1-\left ( \frac{r_+}{r} \right )^{n} \ .
\end{equation}
Again, we assume that in the new solution the $z$ direction is
periodically identified with $z = z + 2\pi R$. This solution has
an event horizon situated at $\rp$ and an ergosurface at $r=\rp
\cosh^{2/n} \beta$, where $\partial_t$ becomes spacelike.

To see quantitatively that this solution carries
both mass and momentum, we calculate the ADM-like stress tensor
for the string with the following asymptotic integrals
\cite{stress}
\be T_{ab}=\frac{1}{16\p G}\oint d\O_{n+1}\,\hat{r}^{n+1}
\,n^{i}\left [\eta_{a b} \left (\partial_i h^{c}_{~c}+\partial_i
h^{j}_{~j} -\partial_{j}h^{j}_{~i} \right ) - \partial_i h_{a b}
\right ]\ . \label{stresstensor} \ee
Here $n^i$ is a radial unit-vector in the transverse subspace and
$h_{\m \n}=g_{\m \n}-\h_{\m \n}$ is the deviation of the
asymptotic metric from flat space. Note that the index labels
$a,b,c\in\{t,z\}$, while $i,j$ run over the transverse directions.
To apply this formula, the asymptotic metric must approach that of
flat space in Cartesian coordinates. This is accomplished with the
coordinate transformation $r=\hat{r}(1+(r_+/\hat{r})^n/2n)$ which
yields
\begin{eqnarray}
ds^2 \simeq - \left (1- \left ( \frac{r_+}{\hat{r}} \right )^n
\cosh^2 \beta \right )dt^2 +2\left ( \frac{r_+}{\hat{r}} \right
)^n \sinh \beta \cosh \beta ~  dt dz  \\
+\left (1 + \left ( \frac{r_+}{\hat{r}} \right )^n \sinh^2 \beta
\right ) dz^2 +\left (1 + \frac{1}{n} \left ( \frac{r_+}{\hat{r}}
\right )^n \right ) dx^i dx_i \ , \nonumber
\end{eqnarray}
keeping only the leading order corrections. Here
$\hat{r}^2=\sum_{i=1}^{n+2}(x^i)^2$.  Hence applying
eq.~\reef{stresstensor}, we find the stress energy for the boosted
black string is:
\begin{eqnarray}
T_{tt}&=&\frac{\Omega_{n+1}}{16\pi G} r_+^n (n \cosh^2 \beta+1 ) \
,
\nonumber\\
T_{tz}&=&\frac{\Omega_{n+1}}{16\pi G}  r_+^n\, n
\cosh\beta\,\sinh\beta \ ,
\label{stressan}\\
T_{zz}&=&\frac{\Omega_{n+1}}{16\pi G} r_+^n (n \sinh^2 \beta-1 ) \
, \nonumber
\end{eqnarray}
where $\Omega_{n+1}$ is the area of a unit $(n+1)$-sphere.
Integrating over $z$, the total energy and momentum of the string
are then
\begin{eqnarray}
E_{\mathrm{bs}} &= & \frac{\Omega_{n+1}R}{8G} r_+^n (n \cosh^2
\beta+1 )
\ , \label{bsenergy} \\
P_{\mathrm{bs} } & = & \frac{\Omega_{n+1}R}{8G}  r_+^n\, n
\cosh\beta\,\sinh\beta \ . \label{bsmomentum}
\end{eqnarray}

The limit of maximal boost $\beta \rightarrow \infty$ results in
divergent $E_{\mathrm{bs}}, P_{\mathrm{bs} }$, but these can be kept
finite if $\rp$ vanishes sufficiently fast.  In particular taking
the large $\beta$ limit while holding $\rp^n \cosh^2 \beta$ fixed
produces finite charges. However, the limiting background has a
naked null singularity at the center of a finite size ergosphere.


\subsection{Comparing Black Strings and Black Holes}
\label{sec:thermodynamics}

Gregory and Laflamme \cite{gla1,gla2} originally gave a simple
argument favoring instability of the static black string by
comparing its entropy to that of a spherical black hole with the
same energy. This argument also plays a role in deducing the full
phase structure of black strings and black holes in a compactified
space-time \cite{kolrev,koltopo}. So we begin here by extending this
discussion of the global thermodynamic stability to the boosted
black string. The analysis for the case at hand becomes slightly
more complicated because, as well as matching the energy, we must
also explicitly match the KK-momentum along the $z$ circle in our
comparison.

We compare the boosted black string solution \reef{eq:metric} to a
$D$-dimensional spherical black hole of radius $\rp'$ moving along
the $z$ axis with velocity $v'=\tanh\beta\,'$. At rest, the energy
of the spherical black hole is \cite{MP}
\[
M_{\mathrm{bh}} =\frac{(n + 2)\Omega_{n + 2}}{16\pi G} {\rp'}^{n + 1}\ .
\]
Now to a distant observer, the spherical black hole behaves like a
point particle and so when boosted, its
energy and momentum are given by
\begin{equation}
E_{\mathrm{bh}} = M_{\mathrm{bh}}\cosh \beta \, '
\ , \quad P_{\mathrm{bh}}= M_{\mathrm{bh}} \sinh
\beta \, ' \ .
\end{equation}
Equating the above to those for the black string given in
eqs.~\reef{bsenergy} and \reef{bsmomentum}, the black hole must
have:
\begin{equation}
\tanh \beta \, ' = \frac{n\cosh\beta\,\sinh \beta}{1+n\cosh^2
\beta}\ , \quad r_+'{}^{n+1} = 2 \pi r_+^n R
\frac{\sqrt{1+n(n+2)\cosh^2 \beta}}{n+2} \frac{
\Omega_{n+1}}{\Omega_{n+2}} \ .\label{house}
\end{equation}
It is interesting to note that with the usual relation
$v=\tanh\beta$, the first expression above can be rewritten as
\begin{equation}
v'=v{n\over n+1-v^2}\ .
\end{equation}
Hence we always have $v'<v$, with $v'$ approaching $v$ (from
below) as $v\rightarrow1$.

We now need to calculate the horizon entropy $S=A/4G$ for each
configuration. For the boosted string, we find
\begin{equation}
S_{\mathrm{bs}} = \frac{\pi R\Omega_{n+1}}{2G} \rp^{n+1}
 \cosh \beta \ . \label{eq:bsentropy}
\end{equation}
The $\cosh\beta$ dependence arises here because proper length
along the $z$ direction at the horizon expands with increasing
$\beta$, as can be seen from eq.~\reef{eq:metric}. In contrast,
the horizon area of the black hole is invariant under boosting.
This invariance is easily verified in the the present case by
explicitly applying a boost along the $z$ direction to the black
hole metric in isotropic coordinates. However, this is a general
result \cite{danny}. Hence for the boosted black hole, we have
\begin{equation}
S_{\mathrm{bh}}  =  \frac{\Omega_{n+2}}{4G} {\rp'}^{n+2}
 \ . \label{eq:bhentropy}
\end{equation}

Setting $S_{\mathrm{bh}}/S_{\mathrm{bs}}=1$ and solving for $R$, we find
\begin{equation}
R_{\mathrm{min}}=\frac{ r_+}{2 \pi \cosh\beta}\,
\frac{(n+2)^{n+2}}{(n(n+2)+\cosh^{-2} \beta)^{n/2+1}}
\frac{\Omega_{n+2}}{\Omega_{n+1} } \ . \label{eq:prediction}
\end{equation}
Hence we might expect that the boosted black string is unstable for
$R>R_{\rm min}$. Fixing $r_+$, $R_{\rm min}$ scales like $1/\cosh
\beta$ for large $\beta$. It should be remembered that the large
$\beta$ limit with $\rp$ fixed has divergent energy. Rescaling $\rp$
while taking the large $\beta$ limit can make the energy finite, but
this causes $R_{\rm min}$ to vanish even more quickly. In any event,
this naive analysis suggests that the instability will persist for
$\beta \rightarrow \infty$. Again, note that the black string
horizon becomes a null singularity in this limit.


\subsection{Instability of Boosted Strings}
\label{sec:boosted_solution_instabilities}

Turning now to the instability of boosted strings, a natural
choice of coordinates in which to perform the analysis are those
for which the string appears at rest:
\begin{equation}
\tilde{t} = \cosh \beta \: t + \sinh \beta \: z \ , \quad
\tilde{z} = \cosh \beta \: z + \sinh \beta \: t \ .
\label{booster}
\end{equation}
In the following we shall refer to this as the ``static frame'', and
our original frame \reef{eq:metric} having simple periodic boundary
conditions in $z$, will be called the ``physical frame''.

Let us begin in the static frame with perturbations having
 functional form $\exp(\tO \tilde{t} + i \tk \tilde{z})$.   Now
transforming back to the physical frame, this becomes $\exp(\Omega
t + i k z)$ where
\begin{equation}
\Omega = \cosh \beta ~ \tO + i \sinh \beta \: \tk \ , \quad k =
\cosh \beta ~ \tk - i \sinh \beta \: \tO \ .
\label{eq:staticfrequencies}
\end{equation}
For real $\tk$ and $\tO$, the imaginary part of $k$ induced by the
boost is inconsistent with the periodic boundary conditions on
$z$, which are imposed in the physical frame. Hence consistency
requires that we add an imaginary part to $\tk$, $i \tanh \beta \,
\tO$, which ensures that the resulting $k$ is real. In practice,
finding solutions also requires adding a small imaginary part to
$\tO$ --- see below. Hence in the static frame, our perturbations
have a $\tilde{t}, \tilde{z}$ dependence of the form
\begin{equation}
\exp[(\tO+i\to) \tilde{t} + i (\tk+ i \tanh \beta \, \tO)
\tilde{z}]\ , \label{eq:replacements}
\end{equation}
where $\tO,\to,\tk$ are all real. In the physical frame, the
$t,z$ dependence then becomes $\exp(\Omega t)\exp\,i(\omega t +k
z)$ where
\begin{equation}
\Omega=\tO/\cosh \beta \ , \quad \omega = \cosh \beta \, \to +
\sinh \beta \ , \tk \quad k = \cosh \beta \, \tk + \sinh \beta \,
\to \ . \label{eq:final_frequencies}
\end{equation}
Again all of the above are real numbers.  Provided we ensure that
$k$ is a multiple of $1/R$, this ansatz is now consistent with the
periodicity of $z$. As before, solutions with $\Omega > 0$ will
correspond to instabilities.

Including the complex part, $i\to$, in the near horizon form of
the solution (\ref{eq:horizonbc}) turns the terms respectively
into in- and out-going modes at the future event horizon. When
$\tO >0$, regularity of the solution requires that we set $B=0$,
as before. For the special case that $\tO$ vanishes, neither
solution diverges on the future event horizon, however, the limit
of the second is undefined there. In this case, we continue to
impose $B=0$ as our boundary condition for $\tO=0$ as this
corresponds to a boundary condition of purely in-going modes at
the future event horizon.

Hence the problem of finding instabilities of the boosted string
reduces to finding instabilities of the static string with the
complex frequencies defined by (\ref{eq:replacements}). With these
frequencies, the perturbations have a time dependent phase.
Boundary condition must therefore be imposed on both the real and
imaginary parts of the unknown function. This means that for each
value of $\tk$ there are two constraints that must be solved on
the horizon, precisely matching the number of free parameters
$\to,\tO$. Apart from these complications, the solutions were
found numerically using the method outlined in Section
\ref{sec:gregory_laflamme_instability}.

\begin{figure}[h]
\begin{center}
\resizebox{7cm}{5cm}{\includegraphics{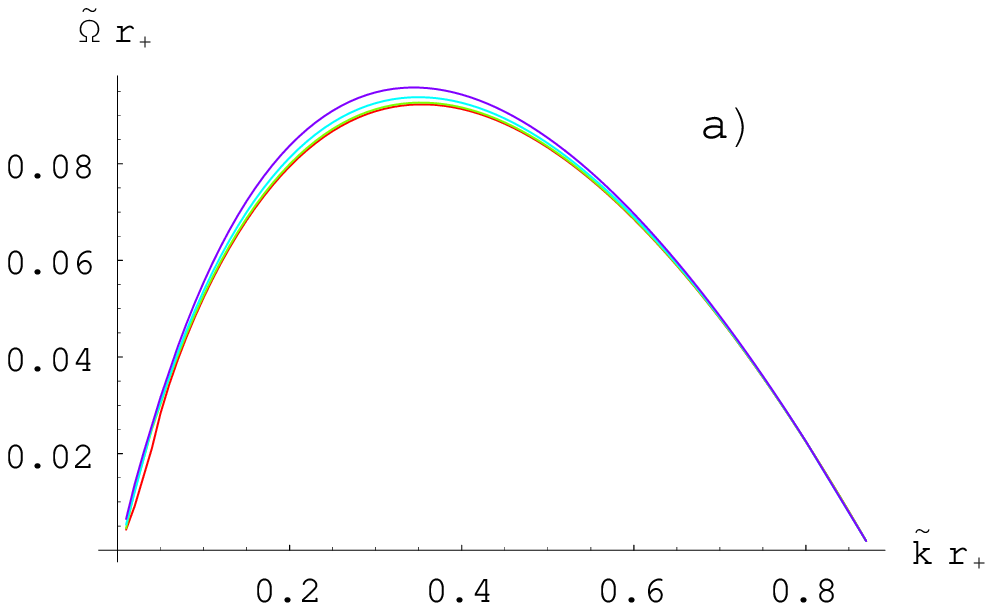}}
\resizebox{7cm}{5cm}{\includegraphics{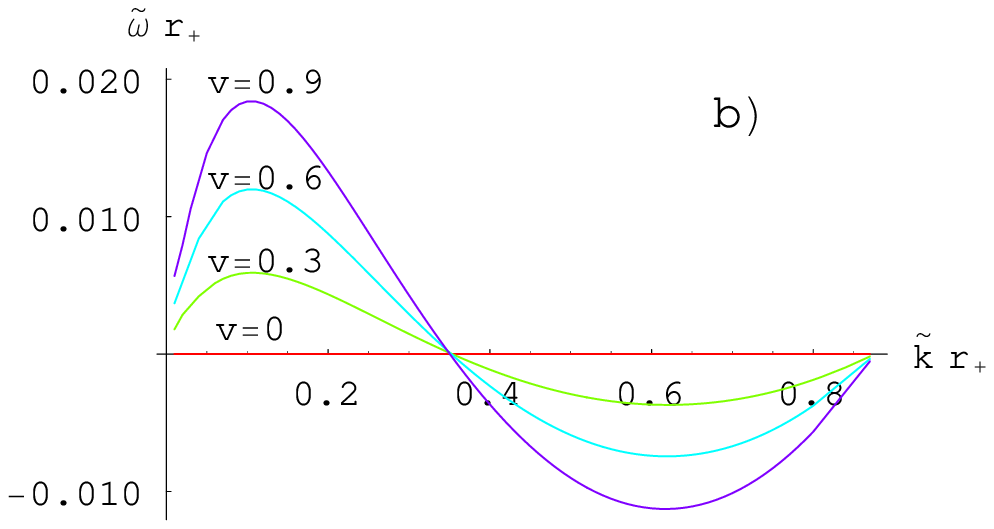}}
\caption{Frequencies $\tilde{\Omega}(\tilde{k})$ and
$\tilde{\omega}(\tilde{k})$ leading to instabilities, as observed
in static ($\tilde{t},\tilde{z}$) frame, for $n=1$. }
\label{fig:solutions_boosted}
\end{center}
\end{figure}

The numerical results for the frequencies $\tO$ and $\to$ in the
static frame are displayed as a function of $\tk$ in Figure
\ref{fig:solutions_boosted} for $n=1$.  The results in other
dimensions are similar.  On the left, we see that $\tO(\tk)$ is
almost independent of the boost velocity $v$. This result might be
interpreted as arising because even when $v=0$, $\tO$ is
suppressed relative to $\tk$ and so making $v$ non-zero (but
small) only yields a small perturbation on the unboosted results.
Further we note that the behavior of $\tO(\tk)$ near $\tk=0$ and
$\tkmax$ is essentially independent of $v$ --- a point we return
to below.

\begin{figure}[h]
\begin{center}
\resizebox{7cm}{5cm}{\includegraphics{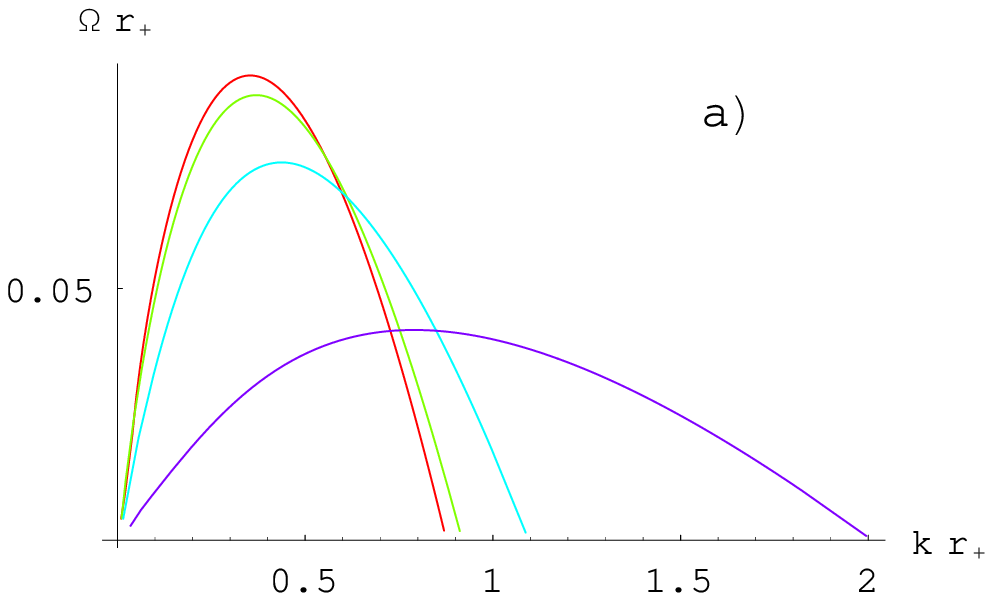}}
\resizebox{7cm}{5cm}{\includegraphics{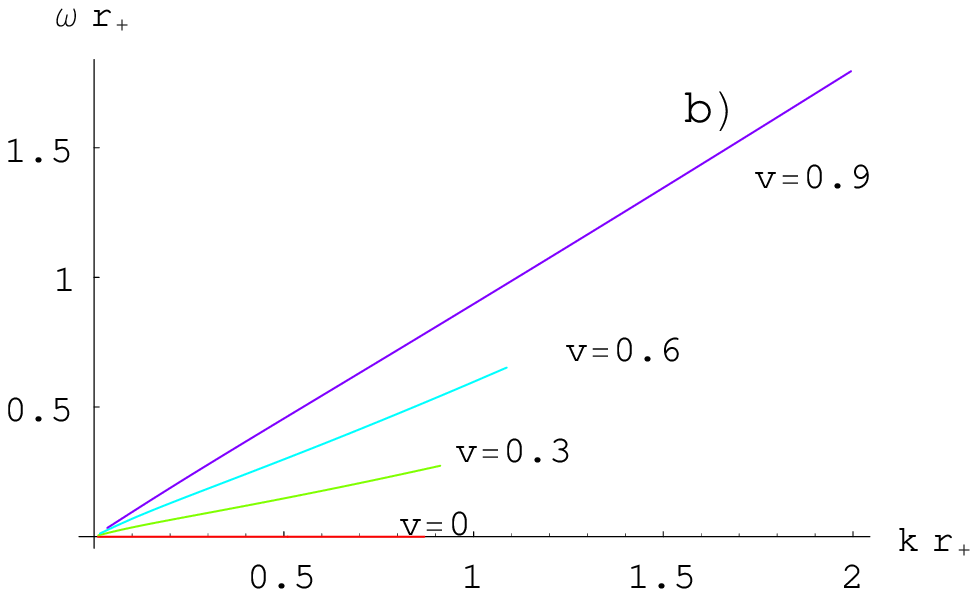}}
\caption{Plot of physical frequencies $\Omega(k)$ and $\omega(k)$
leading to boosted string instabilities for fixed horizon size, at
various boost velocities and with $n=1$. }
\label{fig:plot_vary_v_fixedrp}
\end{center}
\end{figure}

More dramatic differences are seen when the results are transformed
to the the physical frame with eq.~(\ref{eq:final_frequencies}). We
display $\Omega(k)$ in Figure \ref{fig:plot_vary_v_fixedrp}a for
$n=1$. Again, the behavior for other values of $n$ is similar. We
might note that the comparison is made here for boosted strings with
a fixed value of $r_+$. Hence the total energy \reef{bsenergy}
increases as the boost velocity grows and diverges with
$\beta\rightarrow\infty$.

In fact, one can predict the threshold for the Gregory-Laflamme
instability of the boosted string without the numerical analysis
above. The revised ansatz \reef{eq:replacements} in the static
frame was introduced to accommodate the time-dependence of these
modes upon boosting to the physical frame. However, the threshold
mode is defined as that where the time-dependence (in the static
frame) vanishes, \ie $\tO=0$. Hence there is no obstruction to
boosting the threshold mode originally found by Gregory and
Laflamme. Hence there is a simple kinematical relation between the
thresholds for the boosted and static black strings. In the
physical frame, this marginal mode has
\begin{equation}
\kmax = \cosh \beta ~ \tkmax \ , \quad \omega = \sinh \beta ~
\tkmax \label{eq:kcrit_boosted}
\end{equation}
where $\tkmax$ is the threshold for a static black string, listed
in Table \ref{tab:unboosted-coeffs}. Hence these threshold modes
are travelling waves in the $z$ direction having precisely the
same speed as the boosted string.

One may ask whether there are more general modes with $\tO=0$, but
nonzero $\to$. For example an exactly marginal mode in the
physical frame would require that $\tO=0$ and $\to = -\tanh \beta
\: \tk$, but in fact such a solution is inconsistent with the
equations of motion. The linearity of (\ref{eq:hyeqn}) allows us
to arbitrarily choose a normalization in which $h_z$ is real at a
point.  When we set $\tO=0$, the real and imaginary parts of $h_z$
decouple, implying that $h_z$ is real everywhere.  If $\to$ is
non-zero, the only choice of $A$ and $B$ in the near-horizon
solution (\ref{eq:horizonbc}) consistent with $h_z$ real is $A =
B^*$, so that the boundary condition $B=0$ is not possible.  We
then conclude that the only solution with $\tO=0$ is
time-independent in the static frame ($\to=0$), which is then a
travelling wave of constant amplitude in the physical frame.

To close this section, we observe that in the static frame,
$\to(\tk)$ shows some interesting structure, as shown in Figure
\ref{fig:solutions_boosted}b. The zeros of $\to$ seem to be
independent of $v$. The vanishing at $\tkmax$ (and $\tk=0$) is
understood from the discussion above, but there also a fixed
intermediate zero which seems to coincide with the maximum value
of $\tO$. We do not have a physical explanation for the latter.

Using eq.~\reef{eq:final_frequencies}, the phase velocity of the
unstable modes in the physical frame can be written as
\begin{equation}
{\omega\over k}={v+\to/\tk\over 1+v\,\to/\tk}\simeq
v+(1-v^2){\to\over\tk} +\cdots\ . \label{phase-v}
\end{equation}
The last approximation uses our numerical result that generically
$\to/\tk\ll1$. Hence we see that to a good approximation all of
the perturbations travel along the string with the boost velocity
--- a result which is verified by the numerical results in Figure
\ref{fig:plot_vary_v_fixedrp}b. However, given $\to(\tk)$ in Figure
\ref{fig:solutions_boosted}b, we see that the deviations from this
rule are such that the long (short) wavelength modes travel with a
phase velocity that is slightly faster (slower) than $v$. Of course,
the threshold mode moves along the $z$ direction with precisely the
boost velocity.

\subsection{Comparing Black Strings and Black Holes, Again}
\label{sec:thermodynamics2}

\begin{figure}[h]
\begin{center}
\resizebox{8cm}{5cm}{\includegraphics{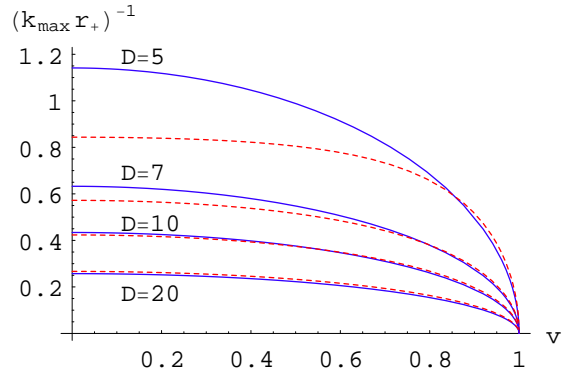}}
\caption{Comparison of the threshold wave-number calculated
numerically \reef{rminbs} (blue, solid) to that predicted by
global entropy considerations (\ref{eq:prediction2}) (red, dashed)
for $D=5,7,10,20$  } \label{fig:kcrit_comparison}
\end{center}
\end{figure}

The threshold mode sets the minimum radius for the compact circle
for which the boosted black string is stable. Hence from
eq.~\reef{eq:kcrit_boosted} above, we have
\begin{equation}
\left({R_{\mathrm{min}}\over r_+}\right)_{\rm BS}={1\over (\kmax
\,r_+)_{\rm BS}} = {1\over\cosh \beta ~ \tkmax\, r_+}
\label{rminbs}
\end{equation}
where again $\tkmax$ is the static string threshold, given in
Table \ref{tab:unboosted-coeffs}. This result might be compared to
that in Section \ref{sec:thermodynamics}. Recall that there we
compared the entropy of the boosted black string to that of a
small black hole boosted along the $z$-direction. In this case, we
found
\begin{equation}
\left({R_{\mathrm{min}}\over r_+}\right)_{\rm BH}={1\over (\kmax\,
r_+)_{\rm BH}} = \frac{1}{2 \pi \cosh\beta}\,
\frac{(n+2)^{n+2}}{(n(n+2)+\cosh^{-2} \beta)^{n/2+1}}
\frac{\Omega_{n+2}}{\Omega_{n+1} } \ . \label{eq:prediction2}
\end{equation}
Hence the simple scaling with $1/\cosh\beta$ in eq.~\reef{rminbs}
is modified here by corrections in powers of $1/\cosh^{2} \beta$.
The two results are plotted together in Figure
\ref{fig:kcrit_comparison} for various spacetime dimensions.

Considering the static results (\ie $v=0$ or $\cosh\beta=1$), Figure
\ref{fig:kcrit_comparison} shows that $(R_{\rm min})_{\rm BS} >
(R_{\rm min})_{\rm BH}$ for smaller values of $D$ but $(R_{\rm
min})_{\rm BS} < (R_{\rm min})_{\rm BH}$ for larger values. Sorkin
\cite{sor} first observed this transition occurs at a critical
dimension between D=13 and D=14.\footnote{The precise value of the
critical dimension may depend on the thermodynamic ensemble
considered \cite{shift}.} This result indicates that there is an
interesting phase diagram \cite{kolrev,koltopo} for $D\le13$, with a
regime $(R_{\rm min})_{\rm BS} > R > (R_{\rm min})_{\rm BH}$ where
the black string is locally stable but the black hole solution is
the global minimum. Further these global considerations then suggest
that in this regime, these two solutions are separated by an
unstable solution describing a non-uniform black string
\cite{wiseman-nonu} --- this structure was recently verified with
numerical calculations for $D=6$ \cite{wiseman2}. In contrast, for
$D>13$, it appears that the non-uniform black string becomes stable
but only appears as the end-state of the decay of the uniform black
string in the regime $(R_{\rm min})_{\rm BH} > R > (R_{\rm
min})_{\rm BS}$ \cite{kolrev}.

\begin{figure}[h]
\begin{center}
\resizebox{8cm}{5cm}{\includegraphics{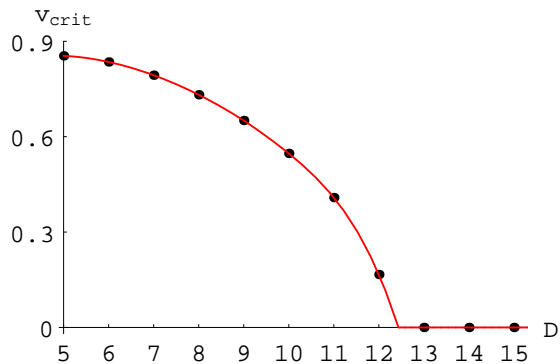}} \caption{The
critical boost at which nonuniform black strings become stable in
various dimensions. (The curve is simply a guide to the eye.)}
\label{fig:vcrit}
\end{center}
\end{figure}

Now we observed that $(R_{\rm min})_{\rm BS}$ and $(R_{\rm
min})_{\rm BH}$ in eqs.~\reef{rminbs} and \reef{eq:prediction2} do
not have the same dependence on the boost velocity. This leads to an
interesting effect which we observe in Figure
\ref{fig:kcrit_comparison}. In the regime $D\le13$, we start with
$(R_{\rm min})_{\rm BS} > (R_{\rm min})_{\rm BH}$ for small
$\cosh\beta$ but there is a transition to $(R_{\rm min})_{\rm BS} <
(R_{\rm min})_{\rm BH}$ for large boosts. Figure \ref{fig:vcrit}
displays the critical boost velocity (for the uniform black strings)
at which this cross-over occurs in various dimensions. This behavior
can also verified using the analytic approximation for the static
threshold provided in \cite{sor}, which yields
\begin{equation}
\left({R_{\mathrm{min}}\over r_+}\right)_{\rm BS}={1\over (\kmax
\,r_+)_{\rm BS}} \simeq {1\over2\pi\cosh
\beta}\left({16\pi\,a\,\gamma^{n+4}\over(n+1)\Omega_{n+1}}\right)^{1/n}
\label{rminbs2}
\end{equation}
where $a\simeq0.47$ and $\gamma\simeq0.686$ are constants.

We note that the minimal radius \reef{eq:prediction2} from the black
hole comparison will receive corrections but these will not change
the final result. The analysis in Section \ref{sec:thermodynamics}
treats the black hole as being spherical sitting inside a fixed
internal circle. For very small black holes this is an acceptable
approximation, but as the size increases, the interactions with the
`image' black holes in the covering space become important and lead
to mass-dependent corrections for the entropy of black holes on
cylinders \cite{cage}. However, as $\beta$ increases, so too does
the proper separation of the black hole and its images (along the
$z$ direction) in their static frame, \ie $\Delta \tilde
z=2\pi\,R\,\cosh\beta'$ where the boost factors are related as in
eq.~\reef{house} but for large boosts,
$\cosh\beta'\simeq\cosh\beta$. Eq.~\reef{house} also shows that the
size of the black holes also grows as $\beta$ increases but at a
much slower rate, \ie $r_+'\propto (\cosh\beta)^{1\over n+1}$ for
large $\cosh\beta$. Hence, the individual black holes effectively
become more and more isolated for large boosts and the approximation
used in Section \ref{sec:thermodynamics} becomes more accurate.

Hence we have reliably found that the critical dimension discovered
in \cite{sor} depends on the boost velocity and in fact disappears
for large values of $\cosh\beta$. Of course, incorporating the
compactification corrections for the black holes \cite{cage}
will allow one to produce a more accurate value for the critical
boost in various dimensions \cite{new}. Further following
\cite{sor}, the critical boost can also be assessed using Gubser's
perturbative construction \cite{gubser-nonu} of the nonuniform black
string \cite{new}.


\section{Black Rings} \label{sec:loop_of_string}
The question of black hole uniqueness in dimension greater than
four was answered decisively by Emparan and Reall with the
construction of an explicit counterexample \cite{ring}. Their
solution is completely regular on and outside a horizon having
topology $S^2 \times S^1$, a black ring.  For the metric, we
consider the form presented in \cite{nonu}:
\begin{eqnarray}
ds^2 & = & - \frac{F(x)}{F(y)} \left ( dt+ R \sqrt{\lambda \nu }
(1+y)  d\psi \right )^2 \label{eq:black_ring} \\
& & + \frac{R^2}{(x-y)^2} \left [
-F(x) \left (G(y)d\psi^2+ \frac{F(y)}{G(y)} dy^2 \right )
+F(y)^2 \left (\frac{dx^2}{G(x)}+\frac{G(x)}{F(x)} d\phi^2 \right )
\right ] \ , \nonumber
\end{eqnarray}
where
\begin{equation}
F(\xi) = 1-\lambda \xi \quad \mbox{and} \quad G(\xi) = (1-\xi^2)(1-\nu \xi) \ .
\end{equation}

Requiring the geometry be free of conic singularities when $F$ or $G$ vanish
determines the periods of the angles $\phi$ and $\psi$ as well as sets
the value of $\lambda$ to one of two possibilities
\begin{equation}
\lambda =
\left \{
\begin{array}{c l}
\frac{2\nu}{1+\nu^2} & \mbox{black ring\,,}\\
1 & \mbox{black hole\,.}
\end{array}
\right . \label{eq:lambda-constraint}
\end{equation}
With the former choice $(x,\phi)$ parameterize a two-sphere while
$\psi$ is a circle.  When $\lambda=1$, $\psi$ joins with $x$ and
$\phi$ to parameterize a three sphere and the solution is a
five-dimensional Myers-Perry black hole \cite{MP} spinning in one
plane.

The family of black ring solution is therefore described by two
free parameters, $\nu$ and $R$.  The first, $\nu$, can be chosen
in the range from $0$ to $1$ and roughly describes the shape of
the black ring.  For $\nu \rightarrow 0$, the ring becomes
increasingly thin and large.  In the opposite limit, $\nu
\rightarrow 1$, the ring flattens along the plane of rotation,
becoming a naked ring-singularity at $\nu=1$.  $R$ can be roughly
thought of as the radius of the ring in a manner that will become
apparent shortly.

\begin{figure}[h]
\begin{center}
\resizebox{7cm}{5cm}{\includegraphics{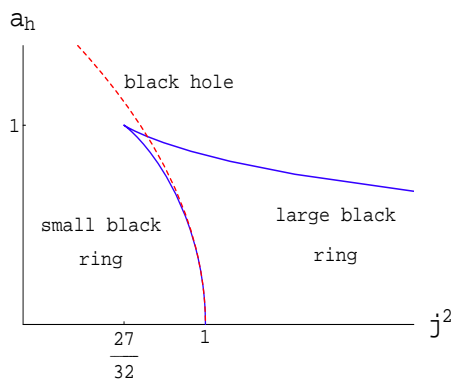}}
\caption{Reduced spin and area for the black ring (blue, solid line)
and black hole (red, dashed) solutions described by the metric
(\ref{eq:black_ring}). The large (small) ring branch corresponds to
$\nu<1/2$ ($\nu>1/2$).} \label{fig:black_ring_phase}
\end{center}
\end{figure}

The ADM energy and spin, as well as the horizon area, are found to
be
\begin{eqnarray}
M & = & \frac{3\pi R^2}{4G} \frac{\lambda(1+\lambda)}{1+\nu}
\ , \label{eq:black_ring_mass} \\
J & = & \frac{\pi R^3}{2G} \frac{(\lambda \nu)^{1/2}
(1+\lambda)^{5/2}}{(1+\nu)^2} \ , \label{eq:black_ring_spin} \\
A & = & 8 \pi^2 R^3 \frac{\lambda^{1/2}(1+\lambda)
(\lambda-\nu)^{3/2}}{(1+\nu)^2(1-\nu)} \ .
\label{eq:black_ring_area}
\end{eqnarray}
A more convenient set of variables for visualizing the various
phases of these solutions are the reduced spin, $j^2$, and area,
$a_{\mathrm{h}}$, defined by
\begin{eqnarray}
j^2 = \frac{27\pi}{32 G}\frac{J^2}{M^{3}}&=&\left \{
\begin{array}{c l}
{(1+\nu)^3\over8\nu} & \mbox{black ring}\\
{2\nu\over 1+\nu} & \mbox{black hole}
\end{array}
\right .\label{redspin}
\\
\qquad a_{\mathrm{h}} = \frac{3}{16} \sqrt{\frac{3}{\pi}}
\frac{A}{(G M)^{3/2}}&=&\left \{
\begin{array}{c l}
2\sqrt{\nu(1-\nu)} & \mbox{black ring}\\
2\sqrt{2{1-\nu\over 1+\nu}} & \mbox{black hole}
\end{array}
\right .\label{redarea}
\end{eqnarray}
We plot the corresponding quantitites in Figure
\ref{fig:black_ring_phase}. Note that the black holes are
described by $a_{\mathrm{h}}=2\sqrt{2(1-j^2)}$. The black rings
lie on two branches, labelled ``large'' and ``small'', which meet
at the critical point $\nu=1/2$.

The ``large'' branch corresponds to solutions where the radius of
the ring grows more quickly than it's thickness, locally
approaching the geometry of a boosted string.  To see this
explicitly, we may take $R \rightarrow \infty$, $\nu \rightarrow
0$ while keeping $R \nu$ fixed. In this limit, we introduce
\cite{nonu}
\begin{eqnarray}
\nu R & = & r_+ \sinh^2 \beta \ , \quad \lambda R =
r_+ \cosh^2 \beta \ , \label{eq:compare}\\
r & = & -R \frac{F(y)}{y} \ , \quad \cos \theta
= x \ , \quad z=R \psi \nonumber \ ,
\end{eqnarray}
and obtain precisely the metric of the boosted black string
(\ref{eq:metric}).  The similarity is in fact more than just
local, comparing the horizon area of the black ring in this limit
we find that it matches the boosted string result
(\ref{eq:bsentropy}), implying that we should indeed take $R$ as a
measure of the radius of the ring.

Given the similarity between boosted black strings and very large
black rings, Emparan and Reall expected that the latter should be
subject to a Gregory-Laflamme type instability \cite{ring}.  Using
(\ref{eq:kcrit_boosted}), the wave-number for the marginal mode of
the five-dimensional boosted string is $\kmax\,r_+\approx 0.876\,
\cosh \beta$. Translating this result to the black ring variables
using (\ref{eq:compare}) yields
\begin{equation}
\kmax \approx \frac{0.876}{R} \frac{\lambda^{1/2}}{(\lambda-\nu)^{3/2}}
 \approx \frac{1.239}{R \nu} \label{eq:compare-kcrit} \ ,
\end{equation}
where the last expression applies only for $\nu\rightarrow0$. Now
$\kmax\gtrsim1/R$ should be the condition for the Gregory-Laflamme
instability to appear in the black ring.\footnote{It is important
here that the unstable mode is localized near the horizon, which is
a point we return to later.} Hence, the above result confirms that
the black ring is unstable in the vicinity of small $\nu$. Further,
considering the second expression above for arbitrary $\nu$, one
finds that $\kmax>1/R$ everywhere which suggests that all of the
black rings are unstable. However, we should not think these
calculations are reliable for all values of $\nu$. We consider this
question in more detail below by studying a simple model of the
black ring.

\subsection{Black Strings to Black Rings}
\label{inging}

Here we would like to construct a simple model of the black ring
that captures its important features. To identify these, we
consider the ratio of the mass and spin of the ring from
eqs.~\reef{eq:black_ring_mass} and \reef{eq:black_ring_spin}. For
small $\nu$, this ratio approaches a constant
\begin{eqnarray}
\frac{MR}{J} & = &  \frac{3}{\sqrt{2}} \left [ 1 - 2\nu
+4\nu^2 + \mathcal{O}(\nu^3) \right ] \nonumber \\
& = & \frac{3}{\sqrt{2}} \left [
    1- 2 \frac{\sqrt{2}GJ}{\pi R^3}
    + 10 \frac{2G^2 J^2}{\pi^2 R^6}
    + \mathcal{O} \left ( \frac{\sqrt{2}G J}{\pi R^3} \right )^3
    \right ] \ , \label{eq:desired_model}
\end{eqnarray}
where implicitly we have expanded the dimensionless quantity
$\sqrt{2}GJ/\pi R^3=\nu+3\nu^2+O(\nu^4)$. Our goal is to reproduce
this expression with a simple string model. So let us assume we have
a spinning loop of string where the loop has a radius $R$ and the
string has a linear ``rest mass'' density $\lambda$. Then we expect,
that up to a boost-dependent factor, the spin is given by $J \sim
\lambda R^2$. This allows us to identify the origin of the most
important contributions to the energy of the black ring, by
re-expressing the contributions in terms of $\lambda$ and $R$.

The constant term in eq.~\reef{eq:desired_model} corresponds to a
contribution to the total energy $\lambda R$, linear in both
factors. Hence remembering to include the boost dependence, this
leading term is simply a combination of the string's rest mass and a
kinetic energy. That this term dominates may have been expected
since we are considering a limit in which the radius of the ring is
large. The next term in the expansion gives a $R$-independent
contribution coming from the gravitational self-energy of the ring
in five-dimensions, $-G\lambda^2$. The final term in
eq.~\reef{eq:desired_model} yields a $1/R$ potential which would
keep the string from shrinking to zero size when formed in a ring.
We can interpret such a contribution as due to rigidity of the
string.

Rigidity has appeared before in various string models. In
particular, it was argued to be necessary to successfully model
the QCD string and was introduced by modifying the Nambu-Goto
action by a term dependent on the extrinsic curvature of the
worldsheet \cite{rigid-string}. It was suggested that such a term
can emerge when the string is constructed as the compactifcation
of a higher-dimensional brane \cite{lindstrom-rigid}.
Compactifying a three-brane on a two-sphere of radius $\rho$ and
forming a loop of string with radius $R$ yields a configuration
where the ratio between the tension and rigidity energies is
$R^2/\rho^2$. Comparing this to the ratio of the first and third
terms in \reef{eq:desired_model} implies that $G \lambda \sim
\rho$, whereas for a boosted black string in five dimensions, we
have $G T_{tt} \sim r_+$ from eq.~\reef{stressan}. This intriguing
coincidence suggests that the rigidity of black strings may be
accomodated by an extension of the ``membrane paradigm''
\cite{membrain} to higher dimensions.

Hence we have argued that the gravitational self-interaction and
rigidity of the black string play a minor role in determining the
configuration for large rings. Now we would like proceed further in
modelling the behavior of such a large black ring by approximating
the latter as a loop of black string and using our results for the
energy and momentum densities of a boosted black string given in
eq.~\reef{stressan}. For a loop of string with radius $R$, these
yield a mass and spin
\begin{eqnarray}
M \equiv 2\pi R\,T_{tt}& = & x\,n\, r_+^n R \left(\cosh
2\beta+1+{2\over n}\right)
\ , \label{eq:model_E}\\
J \equiv 2\pi R^2\,T_{tz}& = & x\,n\, r_+^n R^2 \sinh 2\beta \ ,
\label{eq:model_J}
\end{eqnarray}
where for notational convenience, we have introduced the constant
$x\equiv\Omega_{n+1}/16 G$. Hence we see that our model has three
independent parameters: $R$, $r_+$ and $\beta$, which correspond to
the size and thickness of the loop and the tangential boost velocity
which determines its angular velocity. Given a configuration with
fixed $M$ and $J$, the above equations give two relations between
these parameters but one is left free. Our approach to fixing this
last parameter will be demanding that the ring configure itself to maximize
its entropy:
\begin{equation}
S={A\over 4G} 
=8\pi\,x\,r_+^{n+1}R \cosh \beta \label{eq:model_S}
\end{equation}
This is a straightforward although somewhat tedious exercise. Hence
we only show the salient steps below.

First, we find it useful to replace $R$ by the dimensionless
parameter
\begin{equation}
y\equiv{J\over MR}={\sinh2\beta\over \cosh2\beta+1+{2\over
n}}\label{eq:why}
\end{equation}
where the last expression comes from combining
eqs.~\reef{eq:model_E} and \reef{eq:model_J}. One then determines
$\beta$ and $r_+$ in terms of $y$ as
\begin{eqnarray}
e^{2\beta} &=&{y\left(1+{2\over n}\right)+\sqrt{1+{4\over
n}\left(1+{1\over n}\right)y^2}\over1-y}\ ,\label{eq:beta}\\
r_+^n&=&{M^2\over4xJ}{y\over1+{1\over n}}\left[1+{2\over
n}-\sqrt{1+{4\over n}\left(1+{1\over n}\right)y^2}\right]\
.\label{eq:rplus}
\end{eqnarray}
From these expressions, one can also see that physical solutions are
restricted to the range $0\le y\le1$. Substituting these expressions
into eq.~\reef{eq:model_S} then yields $S(y)$. Plotting the entropy,
one finds that it vanishes\footnote{This vanishing occurs because
$r_+^n$ vanishes at these points, as can be seen in
eq.~\reef{eq:rplus}.} at $y=0$ and 1 and that it has a single
maximum in between. The value of $\ymax$ can be determined
analytically to be:
\begin{equation}
\ymax^2={\sqrt{\left(1+{2\over n}\right)\left(1+{1\over
n}+{1\over4n^2}+{1\over2n^3}\right)}-1+{1\over2n}+{1\over n^2}\over
4\left(1+{1\over n}\right)\left(1+{2\over n}\right)}\label{eq:ymax}
\end{equation}

Now we would like to compare our results to those for the
five-dimensional solution \reef{eq:black_ring}. For $n=1$,
eq.~\reef{eq:ymax} yields $\ymax\simeq.375$ for our loop of black
string while eq.~\reef{eq:desired_model} yields
$y\simeq\sqrt{2}/3\simeq.471$ for the large radius limit of the
exact solution.  Hence our model does not precisely reproduce the
leading result for the large ring, however, the discrepency is only
of the order of $20\%$. Given the simplifying assumptions of our
black string model, it seems to work surprisingly well.

We have found another interesting verification of our model as
follows: In the limit of large $n$, eq.~\reef{eq:ymax} yields
$\ymax^2\simeq1/2n$ and further eqs.~\reef{eq:beta} and
\reef{eq:rplus} indicate that $\beta\simeq1/\sqrt{2n}$ and
$r_+^n\propto1/\sqrt{2n}$, respectively. Hence in this limit (of a
large spacetime dimension), the string loop is very large and thin
while its tangential velocity is small. Therefore it seems
reasonable to treat the loop as a nonrelativistic mechanical string
whose equilibrium configuration can be analysed with Newton's law:
$pv/R=T_{tot}/R$ where the RHS is the centripetal acceleration of a
small element of string with  a linear momentum density $p$ while
the force on the LHS is determined by the total tension. Now
applying a nonrelativistic limit to the stress tensor of the black
string \reef{stressan} yields
\begin{eqnarray}
T_{tz}&=&\rho v\ ,
\label{stressanxx}\\
T_{zz}&=&-T+\rho v^2 \ , \nonumber
\end{eqnarray}
where we distinguish the mass density $\rho$ and the tension $T$ of
the string. For the black string, eq.~\reef{stressan} gives
$T=\rho/n=\Omega_{n+1}r_+^n/16\pi G$ and so we note that we have
$T\ll\rho$ for large $n$, as expected for a nonrelativistic string.
Now setting $p=T_{tz}$ and $T_{tot}=-T_{zz}$, the force law yields
$v^2=T/2\rho=1/2n$ which precisely matches the model result quoted
above.

Hence, it seems that we already have a fairly reliable model of the
black string. Further this model is constructed for an arbitrary
spacetime dimension and so we conclude that black rings also exist
in dimensions higher than five. In fact, for large dimensions, it
seems that a large black ring will be spinning nonrelativistically.

Of course, our simple string model will only capture the leading
behavior of eq.~\reef{eq:desired_model} and not the gravitational or
rigidity corrections. While we do not do so here, one could try
improving our calculations to take these effects into account. In
fact, one indication of the importance of these effects comes from
the black ring solution itself. Note that it has been observed
\cite{nonu} that in the limit of large radius, the five-dimensional
black rings are fairly relativistic in that
$\sinh^2\beta\rightarrow1$, in contrast to our results for large
dimensions above. It is interesting that at this boost corresponds
precisely to that where the tension \reef{stressan} of the
five-dimensional black string vanishes \cite{nonu}, \ie $T_{zz}=0$.
Further, however, looking at \reef{eq:compare} more carefully, we
find
\begin{equation}
\sinh^2\beta={1+\nu^2\over1-\nu^2}\simeq1+2\nu^2\label{toomuch}
\end{equation}
and the black ring actually seems to approach $\sinh^2\beta=1$ from
above as $\nu\rightarrow0$, where the tension of the string would be
negative. Of course, our model only results in a boost where the
black string tension is positive and so can stabilize the spinning
loop. However, the implication of eq.~\reef{toomuch} is that the
stress tensor of black string \reef{stressan} must receive
``rigidity'' corrections, \eg $1/R^2$ terms as in
\cite{rigid-string}, when the string is drawn into a loop so that
the tension remains positive in this limit. Similarly, the
gravitational self-interaction may play a more important role here.

We can also use the black string model to extend our results for the
Gregory-Laflamme instability of boosted black strings to black
rings. In particular, the string loop will be subject to a
Gregory-Laflamme instability when $\kmax R\gsim 1$. Using
eq.~\reef{eq:kcrit_boosted} and Table \ref{tab:unboosted-coeffs}, we
have $\kmax=\cosh \beta ~ \tkmax\simeq .876\,\cosh\beta/r_+$.
Further evaluating these expressions with
eqs.~(\ref{eq:beta}--\ref{eq:ymax}) with $n=1$ gives an instability
for
\begin{equation}
 j^2\gsim.239\ ,\label{answerj}
\end{equation}
where $j^2$ is the reduced spin introduced in eq.~\reef{redspin}.
There we also showed that for the five-dimensional black ring, the
minimum value was $j^2_{\rm min}=27/32\simeq.844$ at $\nu=1/2$.
Hence in accord with the result at the end of the previous section,
these calculations seem to indicate that all of the black ring
solutions will be unstable. However, our model calculations need not
be reliable for small values of $j^2$, \ie for small black rings.

Before addressing the latter question, let us consider a slightly
different approach to evaluating the threshold for the instability
of the black ring. We reconsider our model of a loop of black string
with three independent parameters. As above, we fix the mass and
angular momentum which leaves one free parameter, which we take to
be the radius of the loop. Now rather than extremizing the entropy,
here we require that the proper area of the horizons be the same.
Again this gives three equations determining the model parameters,
$R_{\model},\rp,\beta$, now in terms of the two free parameters of
the black ring $R$ and $\nu$.

This system of equations fixes the rapidity to be
\begin{equation}
\sinh \beta =1\ .\label{gosh}
\end{equation}
It is interesting that this corresponds to the boost for which the
five-dimensional black string becomes tensionless, \ie $T_{zz}=0$,
as is appropriate for the large ring limit. Note here though that we
have not explicitly taken such a limit. The remaining parameters are
found to be
\begin{eqnarray}
R_{\model} & = &  \frac{ (1+\nu)^2 }{1+\nu^2} R \ ,\\
\rp & = & \frac{\sqrt{1-\nu^2} }{1+\nu^2} \nu R \ .
\end{eqnarray}
Note that $R_{\model}$ and $R$ agree in the large ring limit,
$\nu\rightarrow0$, but in general $R_{\model}>R$.

Returning to the Gregory-Laflame instability, the string loop will
suffer from the instability when $\kmax R_{\model}\simeq 1$ with
\begin{equation}
\kmax = \tkmax \cosh \beta = \frac{1.239}{\nu R}
\frac{1+\nu^2}{\sqrt{1-\nu^2}}
\end{equation}
where again we have used the five-dimensional result for $\tkmax$.
Let use consider this threshold more carefully here. The validity of
this model calculation (and that above) requires that the unstable
modes are localized near the horizon on a scale much smaller than
the size of the ring. This is, of course, because our calculations
for the instability of the boosted string assumed an asymptotically
flat metric and so we may only apply these results here if the
perturbation is insensitive to the geometry at the antipodal points
on the ring. Here we are considering the characteristic size of the
modes in the direction orthogonal to the string and hence orthogonal
to the boost direction. Therefore this profile is independent of the
boost velocity and for the threshold mode, we can again use the
results from Section \ref{sec:gregory_laflamme_instability}. The
radial fall-off of this perturbation was determined by the scale:
$\tm = \sqrt{\tO^2+\tkmax^2}=\tkmax$ since $\tO$ vanishes for the
threshold mode. Given the boost factor \reef{gosh} is order one, the
wavelength and the radial extent of the threshold mode are about the
same size.\footnote{Note that we expect the threshold mode has the
least radial extent of the unstable modes.} Hence to be confident of
our calculations for the black ring instability, the estimate above
must be revised to $\kmax R_{\model}\gg 1$, which is equivalent to
\begin{eqnarray}
\frac{\nu \sqrt{1-\nu^2} }{(1+\nu)^2} \ll 1.239\ .
\end{eqnarray}
Notice that the expression on the left-hand side has a maximum of
$0.192$ at $\nu=1/2$ and hence we can be confident that this
inequality will be satisfied in general.

To summarize then, for any black ring on either branch in Figure
\ref{fig:black_ring_phase}, one can find a corresponding black
string model that has the same energy, spin and area. This version
of the calculation again suggests that the black ring are unstable
with a Gregory-Laflamme instability for any value of the parameters.
However, we must note that this calculation is not always reliable.
Recall that our underlying assumption was that the dominant black
ring dynamics were simply determined by the rest energy and tension
of the string. While this is indeed valid for the large black ring
(small $\nu$), eq. \reef{eq:desired_model} clearly shows that this
assumption becomes invalid when $\nu$ grows. In particular, there is
no reason that it should be trusted when $\nu \geq 1/2$ where the
gravitational self-interaction will be important.  For a
conservative bound, we might require that ignoring the gravitational
correction introduces less than a $10\%$ error in the total energy,
which means that we require $\nu \le0.05$. Of course, this bound is
subject to the reader's taste in the admissible error and in any
event, it only represents a bound one's confidence in the validity
of our model. However, these calculations certainly do indicate the
black rings in Figure \ref{fig:black_ring_phase} already experience
a Gregory-Laflamme instability when the reduced spin $j^2$ is of
order one.


\section{Discussion}

We considered the Gregory-Laflamme instability for boosted black
strings. In the static frame, the results are largely unchanged
compared to the instability of a static black string, although the
boundary conditions required a complex frequency with a small
imaginary component. However, the instability is strongly
dependent on the boost velocity in the physical frame, as shown in
Figure \ref{fig:plot_vary_v_fixedrp}a for $n=1$. Since the
threshold mode is by definition time independent, the mode found
for the static black string is also a solution satisfying the
appropriate boundary conditions in the static frame of the boosted
string. As a result for a fixed horizon size, there is a simple
kinematical relation \reef{eq:kcrit_boosted} between the threshold
wave number of the static and boosted black strings. For the
boosted black string, the threshold mode is a travelling wave
moving in the $z$ direction with precisely the same speed as the
boosted string.

In the static case, Sorkin \cite{sor} showed that {\it stable} black
strings and small black holes on a compact circle only coexist below
a critical spacetime dimension, of approximately 13. For the boosted
case, in which there is internal momentum in the circle direction,
we have shown that this critical dimension is boost dependent and in
fact vanishes for large boosts. This result is illustrated in Figure
\ref{fig:kcrit_comparison} by the crossing of the curves for the
minimal radius found from the Gregory-Laflamme analysis and from a
comparison of the entropy of the black holes and strings.

Sorkin's result has interesting implications for the phase diagram
for black objects in a compactified spacetime \cite{kolrev,koltopo}.
For $5\le D\le13$, there is a regime where black holes and stable
black strings coexist. These families of solutions are connected by
a family of unstable and nonuniform black strings. For $D>13$, the
stable black strings and black holes do not coexist and the family
of nonuniform black strings connecting these two phases is now
expected to be stable.

Interest in the nonuniform black strings alluded to above began with
the discussion of \cite{hm1}. Such nonuniform solutions were first
constructed perturbatively by Gubser in five dimensions
\cite{gubser-nonu} and this construction is straightforwardly
extended to any number of spacetime dimensions. Wiseman also used
numerical techniques to find such strings in a fully non-linear
regime in six dimensions \cite{wiseman-nonu}. Here we observe that
these nonuniform strings can be boosted to carry KK-momentum in the
internal direction. First, note that these solutions are static and
periodic in, say, the $\tilde{z}$ direction with period
$2\pi\tilde{R}$. Hence one can compactify these solutions by
imposing the identification:
\begin{equation}
(\tilde{t},\tilde{z})
=(\tilde{t}+2\pi\tilde{R}\tanh\beta,\tilde{z}+2\pi\tilde{R})\ .
\label{idid}
\end{equation}
Now upon boosting as in eq.~\reef{booster}, one arrives in a
boosted frame where the identification is now $(t,z)=(t,z+2\pi R)$
where $R=\tilde{R}/\cosh\beta$. Hence in the physical ($t,z$)
frame, one has a nonuniform string moving with velocity
$\tanh\beta$ along the $z$ direction. Note, however, that we would
not compare nonuniform and uniform black strings with the same
boost factor. As in section \ref{sec:thermodynamics}, any
comparison would fix the total mass and KK-momentum, as well as
the circle radius, and since the ratio of the energy density and
tension of the nonuniform and uniform strings is different so
would be the boost factors for each.

Now our observation on the boost dependence of the critical
dimension has interesting implications for these nonuniform strings.
As in the static case, it would seem that for $D>13$ these strings
are stable for any value of the boost. On the other hand for $5\le
D\le13$, the nonuniform strings would apparently be unstable for low
values of the boost, however, they become stable for large boosts.
Note that in contrast to the uniform string which has a continuum of
unstable modes, the static nonuniform string is expected of have a
single unstable mode below the critical dimension reflecting the
periodicity of the solution \cite{kolrev}. While imposing the
``boosted'' boundary condition \reef{idid} did little to modify the
spectrum of unstable modes for the uniform black string, it seems to
be enough to remove the unstable mode in the nonuniform case. It is
interesting then that these nonuniform boosted black strings may
then form the end state for the decay of the uniform black strings
with KK-momentum.

We also applied our results for the instability of boosted strings
to consider the analogous instability of the black rings of Emparan
and Reall \cite{ring}. Both the naive discussion at
eq.~\reef{eq:compare-kcrit} and the more detailed analysis in
section \ref{inging} seems to indicate that the entire branch of
large-ring solutions is unstable. However, these are both expected
to be reliable for small $\nu$ and so one must limit the application
of our calculations. However, our results certainly indicate that
Gregory-Laflamme instabilities will afflict the black rings already
when the reduced spin $j^2$ is of order one. Hence it seems that
this instability will enforce a Kerr-like bound on this particular
family of solutions. This is then similar to the results of
\cite{ultra} where it was argued that the Gregory-Laflamme
instability played a role in destabilizing ultra-spinning black
holes in $D\ge6$, \ie the only stable spinning black hole solutions
in higher dimensions would have $J^{n+1}\lsim GM^{n+2}$, \ie
$j^2\lsim 1$ for $D=5$. There it was also argued that the
five-dimensional spinning black holes may also become unstable near
$j^2=1$ since there exist ``large'' black rings with the same spin
and mass but a larger horizon area. Recently, it has also been
argued that the small-ring branch is unstable using a thermodynamic
treatment \cite{Arc,Noz}. This result may have been anticipated
since again there are always spinning black holes and large rings
with the same mass and angular momentum but a larger horizon area.

Regarding the internal KK-momentum as a charge, it is interesting to
compare our instability results with those for black strings
carrying a gauge charge \cite{gla2,gla3}, \ie an electric three-form
charge or a magnetic ($n$+1)-form. In common with the gauge charged
string, the maximum value of the growth rate $\Omega$ of the
unstable modes decreases (in the physical frame) as the KK-momentum
is increased, as illustrated in Figure
\ref{fig:plot_vary_v_fixedrp}a. However, one should actually think
of the boosted strings as becoming more unstable as the KK-momentum
grows, since the physical threshold wave number $\kmax$ grows as the
boost factor is increased, as described above. In contrast,
increasing the gauge charge makes the black string more stable by
decreasing the wave number of the threshold mode and it is expected
to be absolutely stable in the extremal limit \cite{gla3}. Note that
the boosted string does not have an extremal limit as
$v\rightarrow1$, rather the horizon becomes a null singularity in
this limit.

We should also contrast our results with those in
\cite{ammw,ross-wiseman,smear}, which consider the Gregory-Laflamme
instability for various black branes in string theory with D0-brane
charge smeared over their worldvolume. In this case, the D0 charge
is introduced by lifting the black brane from ten to eleven
dimensions and boosting in the extra dimension. In contrast to the
present case, there the boost direction and the directions along
which the unstable modes form are orthogonal. In accord with the
discussion here then, the threshold for the boosted solution is
unchanged from that for the original solution, \ie with and without
the D0 charge \cite{ross-wiseman}. Similar boosts of nonuniform
black strings have also been considered to generate new brane
solutions in string theory \cite{what}.

Both $t$ and $z$ remain Killing coordinates for the
gauge-charged strings and it is straightforward to consider boosting
these solutions to form black strings carrying both KK-momentum and
gauge charge. In this case, the threshold for the Gregory-Laflamme
instability would again satisfy the same kinematical relation
\reef{eq:kcrit_boosted} with that for the static string, if we fix
the positions of the inner and outer horizons, $r_\pm$. Hence the
extremal string ($r_+=r_-$) will remain stable even after boosting.
One should note that just as boosting increases the energy density
of the static string, it also increases the gauge-charge density.

The stability of the latter is then relevant for the large radius
limit of the ``dipole-charged'' black rings \cite{emparan-dipole}.
The latter are five-dimensional black rings providing a local source
of an electric three-form charge. This dipole charge is not a
conserved charge and so these solutions introduce an infinite
degeneracy of solutions with the same mass and angular momentum
\cite{emparan-dipole}. Given the above comments, we expect that
introducing a dipole charge on the black rings will make them more
stable. In particular, there should be a family of extremal rings
which are exactly stable for any radius. If one adds further
monopole charges, there also exist supersymmetric black rings
\cite{susy-ring,susy2} which must also be absolutely stable.

The stable dipole-charged rings then include stable solutions where
$J^2/GM^3$ becomes arbitrarily large \cite{emparan-dipole}. Hence
while there is a dynamical Kerr-like bound for the vacuum solutions,
as discussed above, no such bound holds in general. Hence if there
is such a bound in higher dimensions, it must be a more refined
version of Kerr bound, perhaps defined in terms of angular momentum
confined to a finite-size system. Certainly there is no problem
producing configurations with an arbitrarily large (orbital) angular
momentum by taking to slowly moving bodies with very large
separation, even in four dimensions, but, of course, we do not
expect any such Kerr bound to apply to such systems.

While our discussion has focussed on the Gregory-Laflamme
instability affecting black rings, it is possible that these
solutions may suffer from other instabilities as well. For example,
rapidly rotating stars (as modelled by self-gravitating
incompressible fluids) are subject to non-axisymmetric ``bar-mode''
instabilities when the ratio of the kinetic and gravitational
potential energies is sufficiently large \cite{bar-mode}. Given the
discussion of section \ref{inging}, large black rings are certainly
in this regime and so one may suspect that they suffer from a
similar instability. It might be that such instabilities restore the
Kerr bound for black rings with dipole charges but they can not play
this role in general, as again the supersymmetric black rings must
be absolutely stable.\footnote{One can consider non-axisymmetric
deformations of the supersymmetric black rings \cite{susy2} but one
finds that the resulting solutions do not have smooth event horizons
\cite{bald}.}

To consider bar-mode instabilities, one might extend the discussion
of section \ref{inging} to produce a model of the black ring which
is not inherently axisymmetric. The analysis of section
\ref{sec:kksolutions} yields the energy density and tension of a
boosted black string and so one might consider a model in which the
black ring is described by a loop of string with the same mechanical
properties --- this is essentially our model for a uniform spinning
loop. However, this information is insufficient to model general
non-axisymmetric loops. Basically, one still requires an equation of
state for the string. For example, the mechanical string could be
considered a relativistic string characterized by its fundamental
tension plus some internal degrees of freedom. However, there are
many possibilities for the latter, \eg massive or massless
excitations, which would lead to different equations of state but
which could still match the same properties for a uniform boosted
string. Hence progress in this direction requires a greater
understanding of the dynamical properties of the black string.

One of the interesting observations of section \ref{inging} was that
at least in the large-ring limit, the black ring configuration is
essentially determined by the energy density and tension of the
static black string. Hence this invalidates arguments restricting
black rings to five-dimensions based on the interplay of the
gravitational potential and centripetal barrier, which have the same
radial dependence in precisely five dimensions. Rather it shows that
there should be black ring solutions in any number of dimensions
greater than four and it confirms the original intuition presented
in \cite{MP} that the existence of black rings did not depend on the
dimension of the spacetime (as long as $D>4$). Of course, explicitly
constructing these solutions remains a challenging open problem.
Undoubtedly, these are simply one part of the rich multitude of
solutions and physics which remains to be discovered in higher
dimensions.


\section*{Acknowledgments}

It is our pleasure to thank Toby Wiseman for his collaboration in
the early stages of this project.  We also wish to thank \'{O}scar
Dias, Roberto Emparan, Barak Kol, Don Marolf and David Mateos for
discussions and comments. Research at the Perimeter Institute is
supported in part by funds from NSERC of Canada and MEDT of Ontario.
RCM is further supported by an NSERC Discovery grant. JLH was
supported by an NSERC Canada Graduate Scholarship. RCM would also
like to thank the KITP for hospitality in the final stages of
completing this paper. Research at the KITP was supported in part by
the National Science Foundation under Grant No. PHY99-07949.



\end{document}